\documentclass[10pt]{iopart}
\usepackage{iopams} 
\usepackage{setstack}
\usepackage{graphics}
\usepackage{amsthm, graphicx}
\usepackage[colorlinks=true,linkcolor=blue,urlcolor=blue,citecolor=blue]{hyperref}

\newtheorem*{theorem*}{Theorem}
\newtheorem*{question*}{Question}

\newcommand{\bra}[1]{\ensuremath{\left\langle #1\right|}}
\newcommand{\ket}[1]{\ensuremath{\left|#1\right\rangle}}

\begin{document}
\title[Control qubit and nonclassicality in teleportation]{To share and not share a singlet: control qubit and nonclassicality in teleportation}

\author{Kornikar Sen}
\address{Harish-Chandra Research Institute,  A CI of Homi Bhabha National Institute, Chhatnag Road, Jhunsi, Allahabad 211 019, India}
\address{Departamento de Física Teórica, Universidad Complutense, 28040 Madrid, Spain}

\author{Adithi Ajith}
\address{Harish-Chandra Research Institute,  A CI of Homi Bhabha National Institute, Chhatnag Road, Jhunsi, Allahabad 211 019, India}
\address{Indian Institute of Science, CV Raman Road, Bengaluru, Karnataka 560 012, India}

\author{Saronath Halder\footnote{Corresponding author}}
\address{Harish-Chandra Research Institute,  A CI of Homi Bhabha National Institute, Chhatnag Road, Jhunsi, Allahabad 211 019, India}
\address{Center for Theoretical Physics, Polish Academy of Sciences, Aleja Lotnikow 32/46, 02-668 Warsaw, Poland}
\ead{saronath.halder@gmail.com}

\author{Ujjwal Sen}
\address{Harish-Chandra Research Institute,  A CI of Homi Bhabha National Institute, Chhatnag Road, Jhunsi, Allahabad 211 019, India}

\begin{abstract}
The superposition principle provides us the opportunity to unfold many surprising facts. One such fact leads to the generation of entanglement which may allow one to teleport an unknown quantum state from one location to another. We try to understand the role of superposition in the process of quantum teleportation, as a question of potentially fundamental importance. We consider, within the scenario of quantum teleportation, a set-up where the sender and the receiver are in a superposed situation of using a maximally entangled state and not using any entangled state in the teleportation protocol, controlled by a qubit. We address two distinct protocols: in the first case, the sender and the receiver do nothing when they do not have the authority to use entanglement, while in the second case, they still use classical communication even if they do not use entanglement. After accomplishing the protocols, we operate a Hadamard gate on the control qubit, measure the control qubit's state, and consider the outcome corresponding to a particular state of the control. We compare the protocol's fidelity with the maximum fidelity achievable through  
classical resources only. In particular, we provide conditions to achieve nonclassical fidelity in teleportation, in the presence of the control qubit. To explore if there is any quantum advantage (advantage of superposition present in the control qubit), we compare the fidelities of the control qubit-based protocols with the fidelity achieved in a situation where the two parties are in a classical mixture of using and not using the maximally entangled state. We observe that there exists a wide range of parameters defining the initial state of the control qubit for which our protocols provide quantum advantage. To analyse the role of superposition quantitatively, we discuss whether the amount of quantum advantage can be expressed in terms of quantum coherence present in the state of the control qubit.
\end{abstract}
\maketitle

 
\section{Introduction}\label{sec1}
Quantum mechanics allows one to send the information of a quantum state without transporting the actual quantum system. This is possible due to the discovery of quantum teleportation by Bennett \emph{et al}.~\cite{bennett}. After its introduction, quantum teleportation has received tremendous attention (for a review one can have a look into Ref.~\cite{Pirandola} and for other works, the references within that review). Apart from the communicational advantages, there exist various applications and variations of the teleportation protocol, some of which are remote state preparation~\cite{remote}, telecloning~\cite{telecloning}, etc~\cite{adv1,adv2}. One remembers the protocol of entanglement swapping in this respect~\cite{es1,es2,es3,es4}. Teleportation has been experimentally realized using photonic systems shared over free space~\cite{ps1, ps2, ps3, ps4, ps5,ps6}, ion traps~\cite{it1, it3}, nuclear magnetic resonances~\cite{NMR1}, superconducting circuits~\cite{sss1}, optical fibers~\cite{of1,of2,of3}, etc.

Teleportation also plays a significant role in quantum computation \cite{qcar1,qcar2,qcar3,qc1,qc2,qc3,qc4,qc5,qc6}. It is used to integrate various modules of a large quantum processor having a quantum modular structure \cite{qc1,qc2,qc3,qc4}. Moreover, teleportation is also used to send gates between different modules of the quantum processors to perform universal quantum computations \cite{qc5,qc6}.

If two parties share a maximally entangled state, using that resource along with two bits of classical communication from the sender to the receiver, an unknown state of a system can be perfectly teleported from one party to the other~\cite{X1,HHH,Hardy,ujjwal}  (see also~\cite{X2}). If the shared state is partially entangled, the teleportation protocol can still be performed, still providing a quantum advantage, but that can not guarantee perfect teleportation. On the other extreme point, if no entanglement is present between the sender and the receiver, no quantum advantage can be gained from the corresponding teleportation protocol. In the absence of shared entanglement as well as the facility of classical communication, the receiver can just randomly prepare a state which will have the average fidelity $\frac{1}{2}$ to the actual state, in case of qubits. On the other hand, if the sender measures her/his qubit state, classically informs the receiver about its outcome, and the receiver prepares that outcome in her/his lab, then in this measure-prepare process, the maximum fidelity that can be achieved is $\frac{2}{3}$ \cite{1, 2}. In Ref. \cite{ref1}, authors have shown if the fidelity of teleportation is greater than 0.789, the protocol is genuinely quantum. Hence we see entanglement plays a crucial role in teleportation.

To realize the experimental generation of entanglement, readers can take a look into  Refs. \cite{exp1,exp2,exp3,exp4,exp5,exp6,exp7,exp8}.  In Refs. \cite{CQC1,CQC2,CQC3}, the authors have discussed the production of entangled states using cloud-based quantum computers.

Here we ask the following question: what if the sender and the receiver are in a superposition of using and not using a maximally entangled state, with respect to their teleportation capability?

Superposition of entangled and product states have been considered in the literature, and the questions asked there are how much is the output entangled~\cite{new1,new2,new3,new4,new5,new6,new7,new8,new9,new10,new11,new12} and whether the output is entangled~\cite{SU}. 
In this paper, we consider the superposition of two events, one of which {uses} a shared entangled state, while the other does not. Here a question of interest is whether the superposition is going to perform better than a classical mixture of the same events. 
We believe that the present consideration will help us to understand the role of superposition in the process of quantum teleportation.

In~\cite{Jozsa}, Jozsa dealt with a situation where a quantum computer is in a superposition of being switched on and off. It was shown that even in the limit that the quantum computer is almost completely off there is quantum advantage over the corresponding classical scenario. Recently, Siddiqui and Qureshi~\cite{tabish} have considered the double-slit experiment, which has its path detectors in a superposition of being present and absent and in that set-up, it was shown that the wave-particle duality relations are still valid. {Motivated by these results, here we discuss the effect of taking the superposition of a singlet being used in a teleportation protocol or not.} 

A control qubit is a device which can be in a superposition of two orthogonal physical situations, say $\ket{\text{on}}$ and $\ket{\text{off}}$, and correspondingly the transformation of a system may be controlled by the state of the qubit. The action of the control qubit has often been discussed in the context of quantum switch introduced in the context of indefinite causal orders~\cite{chiribela1,chiribela2,ICO3,ICO4,ICO6,ICO8,qmetro,qcom2,qcom3,qcom4,chiranjib}. Applications of quantum switches provide advantages in many quantum tasks, and for example, it helps to increase the precision of quantum metrology~\cite{qmetro}, diminishes complications of quantum communications~\cite{qcom1,qcom2,qcom3,qcom4}, etc. The concept of quantum switch has also been studied in the context of teleportation protocols \cite{chiranjib, CVT}. In~\cite{chiranjib}, the authors have considered two teleportation channels in superposition of causal orders controlled by a quantum switch and have shown that even if noisy mixed states are used as resource, much higher fidelity can be achieved than the classical case. Applications of quantum switches in continuous-variable teleportations are discussed in~\cite{CVT}. To explore experiments on quantum switches, see Refs.~\cite{switch_exp2,switch_exp3,switch_exp4,switch_exp5,switch_exp6}.

To describe the superposition of the two distinct situations, we introduce a control qubit which dictates the {usage} of the shared entanglement in the teleportation protocol. We represent the scenario where the maximally entangled state is {being used} by considering the control qubit to be in the $\ket{\text{on}}$ state, and in parallel, the $\ket{\text{off}}$ state of the same qubit represents the situation when Alice and Bob do not {use} any shared entanglement. 

The initial state of the control qubit is chosen to be any arbitrary pure state, so that it does not share any classical or quantum correlation with Alice's or/and Bob's systems. When the control qubit is in the $\ket{\text{off}}$ state, there can at least be two possibilities: Alice and Bob can do nothing, or they can use classical communication to send information about the qubit to be teleported. It is to be noted that  they were also allowed to use classical communication when the control was in $\ket{\text{on}}$. Correspondingly, we define two protocols: in the first one (Protocol 1), they sit idle {when the control is in $\ket{\text{off}}$ state}, and in the second protocol (Protocol 2), they are allowed to use classical communication even when {they do not have the facility to use the shared maximally entangled state}. In both protocols, when the control qubit is in the $\ket{\text{on}}$ state, the usual teleportation is performed. Since the additional qubit introduced in this work controls the actions of Alice and Bob, the protocols may create entanglement between the control qubit and the rest of the system, but no additional entanglement will be created between Alice and Bob. To realise the effect of quantumness in a superposition of the two situations, after performing the protocols, we rotate the control qubit's state using a Hadamard gate on it. Finally, we probabilistically project the control qubit's state on $\ket{\text{on}}$. We quantify the performance of the complete process through the fidelity between the desired state to be teleported and the actual received state.

We observe that in the case of Protocol 1, there exist some regions where the fidelity becomes worse than $\frac{1}{2}$. But in Protocol 2, though there exist situations where the fidelity may become less than $\frac{2}{3}$ or $0.789$, it always remains better than $\frac{1}{2}$, which is expected since in the second protocol, the performers do make use of classical communication. We also compare the protocols with the respective situations where there is no control qubit and the two parties {use} a maximally entangled state with a certain probability, that is, instead of superpositions we consider classical mixtures of the two situations. We prove that the phase between the $\ket{\text{on}}$ and $\ket{\text{off}}$ states, and not the probability amplitudes, is enough to detect whether there will at all be any advantage of using control qubit over the classical mixture. We refer to it as a quantum advantage. Furthermore, we establish a relation between the quantum advantage and quantum coherence of the initial control qubit's state. It is proved that in a particular range of parameters, the advantage is a monotonically increasing function of quantum coherence of the control qubit, measured in the computational basis. Because of computational simplicity, we have analyzed single-qubit teleportation only, but the present idea can also be generalized to higher dimensions. 

Though the idea of taking superposition of the maximally entangled state being used and not used is motivated by the work of Jozsa \cite{Jozsa}, in our case, we find the teleportation fidelity is not nonclassical in the limit in which the shared entanglement is not being used. Therefore, even though the motivation is taken from the very interesting work of Jozsa, it does not immediately infer the result that is obtained in the current case.

The rest of the paper is organized as follows: in Sec.~\ref{sec2}, we give a brief description of the standard teleportation protocol and about a small modification that we add. The exact scenario which we examine is introduced in Sec.~\ref{sec3}. The discussions and results of Protocol 1 are also presented in the same section. We move to Protocol 2 in the next section, i.e., in Sec.~\ref{sec4}. The relation of quantum coherence with the quantum advantage is explored in Sec.~\ref{sec5}. Finally, our concluding remarks are presented in Sec.~\ref{sec6}.

\section{Prerequisites}\label{sec2}
In a teleportation protocol, a sender, Alice, tries to teleport the state of an unknown qubit $A'$ to a receiver, Bob, using local quantum operations and classical communication (LOCC) instead of physically sending the quantum system, but using pre-shared entanglement. Let the state of $A'$ be $\ket{\psi}_{A'}$. As a resource of the protocol, Alice's lab contains another qubit, $A$, which is entangled with a qubit of Bob's lab denoted by $B$.

The exact teleportation of the state of $A'$ requires the qubit-pair $AB$ to be in a maximally entangled state, say the singlet, $\ket{\psi^-}_{AB}=\frac{1}{\sqrt{2}}(\ket{01}-\ket{10})$. The steps of the actual protocol are described below:

\begin{enumerate}
\item Alice jointly measures the state $\ket{\psi}_{A'}$ and her part of the state $\ket{\psi^-}_{AB}$ in the Bell basis, $\{\ket{\phi^+}_{A'A},\ket{\phi^-}_{A'A},\ket{\psi^+}_{A'A},\ket{\psi^-}_{A'A}\}$, where $\ket{\phi^\pm}_{A'A}=\frac{1}{\sqrt{2}}(\ket{00}\pm\ket{11})$ and $\ket{\psi^\pm}_{A'A}=\frac{1}{\sqrt{2}}(\ket{01}\pm\ket{10})$.
    
\item Using classical communication, Alice informs the output of her measurement to Bob.
   
\item Depending on the output, Bob applies one of the operators, $\{\mathcal{I}_B,\sigma_x,\sigma_y,\sigma_z\}$ on his qubit. Here $\mathcal{I}_B$ denotes identity operator on the Hilbert space representing Bob's qubit and $\sigma_x$, $\sigma_y$, $\sigma_z$ are Pauli matrices acting on the same Hilbert space.
\end{enumerate}

Along with the above steps from the original teleportation protocol, we add another step.

\begin{itemize}
\item Alice, at the end, applies one of the operators $\{\mathcal{I}_A,\sigma_x,\sigma_y,\sigma_z\}$ on the qubit $A$, depending on the result of her measurement, to transform her two-qubit state into a singlet. Here the operators carry the same meaning as above, but defined to act on a qubit Hilbert space of Alice, viz. $A$, and for example, $\mathcal{I}_A$ is again an identity operator but acts on the Hilbert space of $A$. 
\end{itemize}

The usual teleportation protocol does not involve this additional step. We have included this step to get back to the state we started from, that is, to complete the round. Since our analysis mostly focuses on the fidelity between the state received by Bob and the initial state that Alice wanted to teleport, the results would not change significantly if the additional step is excluded.

Though the teleportation protocol involves classical communication, the complete process of the protocol can be described using the following Kraus operators:

\begin{eqnarray*}
&&K_1=(\ket{\psi^-}\bra{\psi^-})_{A'A}\otimes \mathcal{I}_B,\\
&&K_2=(\mathcal{I}_{A'}\otimes (\sigma_z)_{A}\otimes\mathcal{I}_B)((\ket{\psi^+}\bra{\psi^+})_{A'A}\otimes(\sigma_z)_B),\\
&&K_3=(\mathcal{I}_{A'}\otimes (\sigma_x)_{A}\otimes\mathcal{I}_B)((\ket{\phi^-}\bra{\phi^-})_{A'A}\otimes(\sigma_x)_B),\\
&&K_4=(\mathcal{I}_{A'}\otimes (\sigma_y)_{A}\otimes\mathcal{I}_B)((\ket{\phi^+}\bra{\phi^+})_{A'A}\otimes(\sigma_y)_B),
\end{eqnarray*}
where $\mathcal{I}_{A'}$ represents the identity operator acting on the qubit $A'$. The output state of the protocol is

\begin{equation*}
\rho'=\sum_{\mu=1}^4 K_\mu \rho K_\mu^\dagger=(\ket{\psi^-}\bra{\psi^-})_{A'A}\otimes(\ket{\psi}\bra{\psi})_B,
\end{equation*}
where $\rho=(\ket{\psi}\bra{\psi})_{A'}\otimes(\ket{\psi^-}\bra{\psi^-})_{AB}$
is the density matrix representation of the initial state (before the teleportation protocol is performed). The operators, $K_\mu$ $\forall \mu = 1,2,3,4$, are defined in such a way that $K_\mu \rho K_\mu^\dagger=\frac{1}{4}(\ket{\psi^-}\bra{\psi^-})_{A'A}(\ket{\psi}\bra{\psi})_B$ $\forall \mu$.

The teleportation protocol can be described without using the apparatus of Kraus representation. Here we have used the Kraus operator representation because in the next part, where we will introduce a control qubit, the new protocol can be efficiently formulated in terms of Kraus operators. The set of Kraus operators considered here is not unique, any other set of Kraus operators, the action of which represents teleportation protocol, could have been taken. We think that consideration of any other suitable set of Kraus operators would not change the results qualitatively.

If instead of pursuing the teleportation protocol, Bob randomly guesses the state and creates it in his lab, then the average fidelity of the randomly generated state to the actual state is $\frac{1}{2}$. On the other hand, if the maximally entangled state is not available but utilization of the facility of classical communication is still possible, then the maximum average fidelity can be raised to $\frac{2}{3}$ using the measure-prepare protocol \cite{1,2}. In this protocol, Alice measures the state that she wants to teleport and informs the measurement outcome to Bob through classical communication. Bob, after knowing the measurement outcome from Alice prepares his qubit in the same state as the measurement outcome. In Ref. \cite{ref1}, the authors have shown the fidelity of a teleportation protocol should be more than 0.789 to certify the genuine quantumness of the protocol. We want to compare our protocol with these fidelities, identifying the situations where lower than these fidelities are obtained, as instances of separate ``classical'' scenarios.

\section{No classical communication when the control is off}\label{sec3}
In this paper, we consider the teleportation protocol to be controlled by a third party, which is another two-level quantum system. The levels of this control qubit indicate if {the singlet, shared between Alice and Bob, is being used in the protocol or not}. When we say `the control qubit is on' it means the singlet is {allowed to be used}, and the corresponding state of the control qubit is denoted as $\ket{\text{on}}$, similarly, when `the control qubit is off' the state is represented as $\ket{\text{off}}$ and it indicates that Alice and Bob {do not have the permission to use the singlet}. Any two orthogonal states can be used in place of the $\{|\text{on}\rangle,|\text{off}\rangle\}$ states of the control qubit. For example the two polarisation states of photons can be a possible choice. 
There are various works where photons are created in superposition of orthogonal states, and using such a state as a control qubit, gates are operated on the system in different orders to produce indefinite causal order (see Refs. \cite{switch_exp2,switch_exp3,switch_exp4,switch_exp5,switch_exp6}). We hope that using similar methods, our protocol can also be efficiently implemented experimentally. Here, in place of considering different orders of the same gates, the experimentalist  needs to apply a “quantum gate” and a “classical gate” on the system depending on the control qubit’s state, which in this case can be any photon state. By quantum or classical gate we mean the gates which implement the teleportation using quantum correlation and classical correlation, respectively. 

In short, we are interested in the following two physical situations:
\begin{eqnarray*}
&& |\text{on}\rangle_C |\psi\rangle_{A'} |\psi^-\rangle_{AB}~~ \mbox{and} \label{achhi}\\ 
&&|\text{off}\rangle_C |\psi\rangle_{A'}|\psi^-\rangle_{AB}. \label{nei}
\end{eqnarray*}
When the control qubit is in the state \(|\text{on}\rangle\), we follow the usual teleportation protocol. Thus, the state of the four qubits including the control's state transforms as
\begin{equation*}
|\text{on}\rangle_C |\psi\rangle_{A'} |\psi^-\rangle_{AB} \rightarrow |\text{on}\rangle_C |\psi^-\rangle_{A'A} |\psi\rangle_{B}.
\end{equation*}
Since the $\ket{\text{off}}$ state of the control qubit implies the {non-usage} of shared entanglement between Alice and Bob, in such a situation, no quantum resource can be {utilized} for teleportation. In this section, we consider the scenario for which, when the control qubit is in $\ket{\text{off}}$ state, the parties will not even have any classical commuication channel. Thus the corresponding transformation is just the identity, and is given as
\begin{equation*}
|\text{off}\rangle_C |\psi\rangle_{A'} |\psi^-\rangle_{AB} \rightarrow |\text{off} \rangle_C |\psi\rangle_{A'} |\psi^-\rangle_{AB}.
\end{equation*}

Our next aim is to mathematically formulate a process, where, when the control qubit is in the $\ket{\text{on}}$ state, the ordinary teleportation protocol using a maximally entangled state would be performed, and when the control qubit is in the $\ket{\text{off}}$ state, a classical protocol would be followed to teleport the state without using any shared quantum correlation. To describe this operation of teleportation of the state of Alice’s qubit to Bob depending on the state of the control qubit, we define the the following Kraus operators:
 \begin{equation}
M_\mu=\ket{\text{on}}\bra{\text{on}}\otimes K_\mu+\ket{\text{off}}\bra{\text{off}}\otimes\frac{1}{2}\mathcal{I}_{A'AB}, \label{eq1}
 \end{equation}
where $\mu=$1, 2, 3, 4 and the identity operator, $\mathcal{I}_{A'AB}$, acts on the composite Hilbert space consisting of Alice's two qubits, $A'$ and $A$, and Bob's qubit, $B$. $\mathcal{I}_{AA'B}$ has been pre-factored with a 1/2 so that the relation $\sum_{\mu=1}^4 M_\mu^\dagger M_\mu=\mathcal{I}_{CA'AB}$ is satisfied. The operators, $M_\mu$, act on the composite Hilbert space consisting of the control qubit, $C$, along with $A'$, $A$, and $B$. 
One can notice from the form of the Kraus operators that when the state of the control qubit is $\ket{\text{on}}$, the Kraus operators representing the quantum teleportation protocol (where the maximally entangled state will be used) will be applied, and when the control qubit is in the $\ket{\text{off}}$ state, the classical teleportation (when no entanglement will be used) protocol’s Kraus operators will be operated on the state of Alice and Bob’s qubits. To define a superposition of these two events, we have taken the linear combination of the two sets of Kraus operators.

This kind of structure of the Kraus operators can also be seen in the works on quantum switches where two or more definite causal orders of operations are superposed to create an indefinite causal order of operations~\cite{chiribela1,chiribela2,ICO3,ICO4,ICO6,ICO8,qmetro,qcom2,qcom3,qcom4,chiranjib}. Though here we are not superposing two causal orders, we are still superposing two distinct operations. Therefore, the structure of the Kraus operators defined here is motivated by the ones considered in the works of quantum switches. 

The Kraus operators of Eq.~(\ref{eq1}) represent the superposition of two LOCC operations between Alice and Bob. Thus the protocol introduced through these Kraus operators is unable to create or increase entanglement between Alice and Bob. The protocol, if possible, can only use the quantumness of the control qubit as an additional resource.

We would like to bring the attention of the readers to the fact that the control qubit introduced in the protocol is not being measured initially. The Kraus operators, $M_\mu$, are defined in such a way that they, as a whole, read the control qubit’s state and decide to teleport if it is in an “on” state. Just like a shift operator used in quantum walks where, for example, the operator shifts the particle one step towards right or left depending on the state of the coin operator, without measuring the coin operator’s state. 

As it was mentioned earlier, the system controlling the availability of the singlet involves quantumness. A state of the control qubit, say $\alpha\ket{\text{on}}+\beta\ket{\text{off}}$,  represents a suporposition of two distinct situation, viz., `to {use}' or `not {use}' entanglement, where $|\alpha|^2+|\beta|^2=1$. We consider the initial joint state of the four qubits, participating in the teleportation protocol, as
 \begin{equation*}
\ket{\xi}=(\alpha|\text{on}\rangle+ \beta|\text{off}\rangle)_C |\psi\rangle_{A'}|\psi^-\rangle_{AB}. 
\end{equation*}

Since we have considered an additional object, i.e., the control qubit, the introduced protocol consists of a larger set-up compared to the usual teleportation protocol. Moreover, the manipulation and measurements performed on the control qubit may require additional resources. But we would like to mention that the initial state considered in this protocol still consists of one ebit of entanglement between the sender and the receiver which is the same as the amount required for perfect teleportation of one-qubit of information, following usual teleportation protocol. 

By acting the Kraus operators on the initial state, $\ket{\xi}$, we get the final state after performance of Protocol 1, given by  

\begin{eqnarray*}
&&\rho^{\text{Pr}_1}_{CA'AB}=\sum_{\mu=1}^4 M_\mu \ket{\xi}\bra{\xi} M_\mu^\dagger\\
   && =|\alpha|^2\ket{\text{on}}\bra{\text{on}}\otimes \sum_{\mu=1}^4 K_\mu\rho K_\mu^\dagger+\frac{\alpha\beta^*}{2}\ket{\text{on}}\bra{\text{off}}\otimes\sum_{\mu=1}^4 K_\mu\rho\\&&+\frac{\alpha^*\beta}{2}\ket{\text{off}}\bra{\text{on}}\otimes\sum_{\mu=1}^4 \rho K_\mu^\dagger+|\beta|^2\ket{\text{off}}\bra{\text{off}}\otimes \rho.
\end{eqnarray*}

We have considered the state of the control qubit to be normalized, i.e., $|\alpha|^2+|\beta|^2=1$. Hence omitting the global phase, without loss of generality, we can consider $\alpha=\cos\frac{\theta}{2}$ and $\beta=e^{-i\phi}\sin\frac{\theta}{2}$, where $\phi\in[0,2\pi)$ and $\theta\in[0,\pi]$.

After getting the state $\sum_\mu M_\mu \ket{\xi}\bra{\xi} M_\mu^\dagger$, we rotate the state of the control qubit by a Hadamard operator, $H_s$. $H_s$ transforms \(|\text{on}\rangle\) and \(|\text{off}\rangle\) to \((|\text{on}\rangle + |\text{off}\rangle)/\sqrt{2}\) and \((|\text{on}\rangle - |\text{off}\rangle)/\sqrt{2}\) respectively. Hence the form of the final state is $H_s\otimes I_{A'AB} \left(\rho^{\text{Pr}_1}_{CA'AB}\right) H_s^\dagger\otimes I_{A'AB}$. After the application of the operator, we measure the control qubit's state in the \(\{|\text{on}\rangle, |\text{off}\rangle\}\) basis and choose the whole output when the outcome is associated with the \(|\text{on}\rangle\) state (briefly, we say this as `the outcome is the $\ket{\text{on}}$ state'). When a qubit state, say $\eta$, is measured on a qubit basis, say $\{|\gamma\rangle,|\gamma^\perp\rangle\}$, the probability that the state would be projected on $|\gamma\rangle$ ($|\gamma^\perp\rangle$) is $p(\gamma)=\langle \gamma |\eta|\gamma\rangle$ $\left(p(\gamma^\perp)=\langle \gamma^\perp |\eta|\gamma^\perp\rangle\right)$. Hence when we measure the control qubit in $\{|\text{on}\rangle, |\text{off}\rangle\}$ basis, the probability of getting $|\text{on}\rangle$ as output is $\text{tr}\left(\langle \text{on}|H_s\otimes I_{A'AB} \left(\rho^{\text{Pr}_1}_{CA'AB}\right) H_s^\dagger\otimes I_{A'AB}|\text{on}\rangle\right)$, which we find to be equal to $(4-\sin\theta\cos\phi)/8$. The final nonnormalized output state, following this path, is given by
\begin{equation*}
    \rho^{\text{Pr}_1}_{A'AB}=\sum_{\mu=1}^4 \left(\frac{|\alpha|^2}{2}K_\mu\rho K_\mu^\dagger+\frac{\alpha\beta^*}{4}K_\mu\rho+\frac{\alpha^*\beta}{4}\rho K_\mu^\dagger\right)+\frac{|\beta|^2}{2}\rho.
\end{equation*}
 After normalizing the state and taking trace over $A$ and $A'$ we can obtain the state of the qubit on Bob's side, which is given by
\begin{equation*}
   \rho^{\text{Pr}_1}_{B}=\frac{4|\alpha|^2-\alpha\beta^*-\beta\alpha^*}{4-\alpha\beta^*-\beta\alpha^*}\ket{\psi}\bra{\psi}+\frac{2|\beta|^2}{4-\alpha\beta^*-\beta\alpha^*}\mathcal{I}_B.
\end{equation*}
Hence the fidelity of the state of $B$ to the initial state of $A'$ in terms of $\theta$ and $\phi$ is given by
\begin{eqnarray}
    \mathcal{F}_{\text{Pr}_1}(\theta,\phi)&=&\frac{2(1+\cos^2{\frac{\theta}{2}})-\sin{\theta}\cos{\phi}}{4-\sin{\theta}\cos{\phi}} \nonumber\\
    &=& \frac{3+\cos\theta-\sin\theta\cos\phi}{4-\sin\theta\cos\phi},\label{eq2}
    \end{eqnarray}
which is independent of $\ket{\psi}_{A'}$. 
Instead of $\ket{\text{on}}$ state, if the output corresponding to $\ket{\text{off}}$ state is chosen, then the corresponding fidelity is given by- $\mathcal{F'}_{\text{Pr}_1}=\frac{3+\cos\theta+\sin\theta\cos\phi}{4+\sin\theta\cos\phi}$. Thus, we see the fidelities corresponding to the measurement outcomes $\ket{\text{on}}$ and $\ket{\text{off}}$ are interchangeable under the transformation $\phi\rightarrow\phi+\pi$. However, unless specified, we mostly discuss about $\mathcal{F}_{\text{Pr}_1}$. 

In the ideal teleportation protocol using a maximally entangled state, the desired state can be perfectly teleported, achieving unit fidelity. The motivation in this work is not to reach that fidelity, which is of course not possible since there are situations when we are not using the entanglement resource. We first want to compare this protocol with the classical situation with no shared entanglement. To examine the effect of the ``quantumness" contained in the control, in the next part, we will compare it with the situation where the control qubit is in a classical state, representing the classical mixture of the two situations that are {teleportation by using} and not {using} entanglement.

\begin{figure}[h!]
\includegraphics[scale = 1.0]{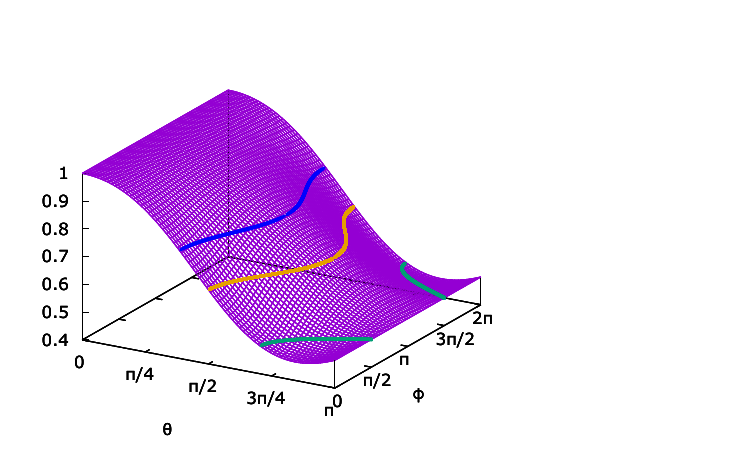}
\caption{Fidelity of teleported state to the actual state in Protocol 1. We plot $\mathcal{F}_{\text{Pr}_1}$ along the vertical axis as a function of the control qubit-state's parameters $\theta$ and $\phi$ which are represented in radians in the horizontal axes. The vertical axis is dimensionless. The functional form of fidelity plotted using the violet surface is expressed in Eq.~(\ref{eq2}). To compare the fidelity of Protocol 1 with the fidelity of random guess, with the maximum fidelity accomplished through the measure-prepare classical process without shared entanglement, and with the bound on the fidelity beyond which it confirms genuine quantumness of the teleportation protocol, we present the curves along which the planes, $\mathcal{F}_{\text{Pr}_1}=\frac{1}{2}$, $\mathcal{F}_{\text{Pr}_1}=\frac{2}{3}$, and $\mathcal{F}_{\text{Pr}_1}=0.789.$ intersect the violet surface using green, sky-blue, and orange colors, respectively.}\label{fig1} 
\end{figure}

\begin{figure*}[t!]
\includegraphics[scale=.43]{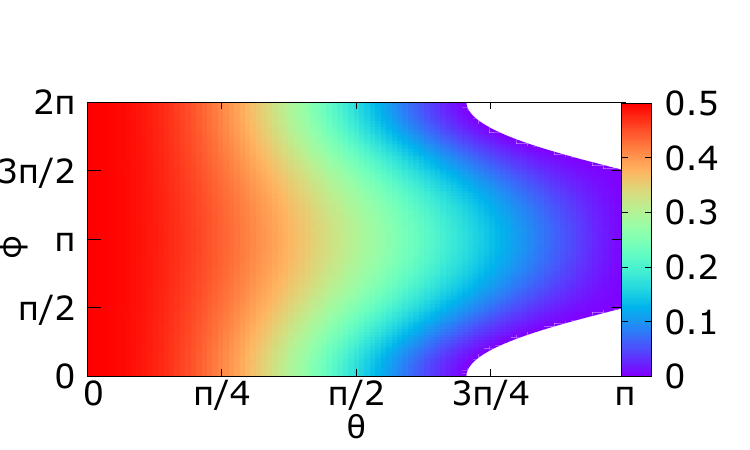}
\includegraphics[scale=.43]{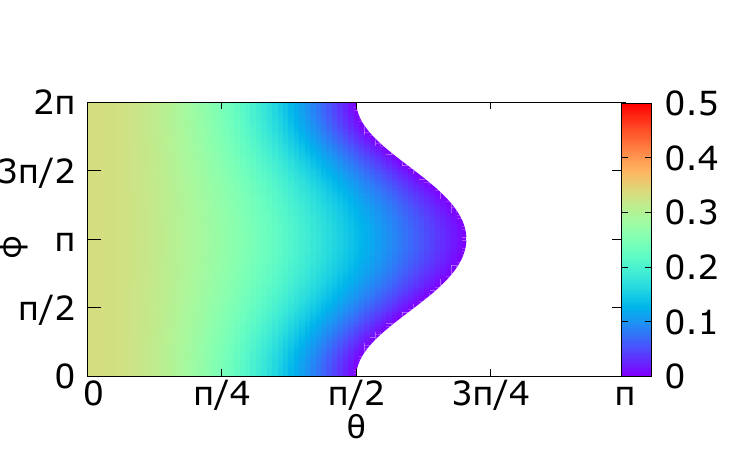}
\includegraphics[scale=.43]{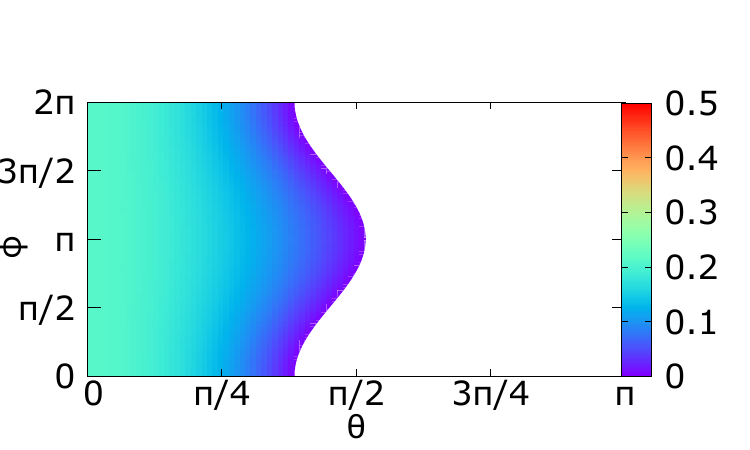}
\caption{Advantages of the quantum teleportation protocol over random guess and classical protocols. We plot $\mathcal{F}_{\text{Pr}_1}-\frac{1}{2}$ (left panel), $  \mathcal{F}_{\text{Pr}_1}-\frac{2}{3}$ (middle panel), and $  \mathcal{F}_{\text{Pr}_1}-0.789$ (right panel) as functions of the angles $\theta$ and $\phi$. The horizontal and vertical axes represent the parameters $\theta$ and $\phi$ respectively and are presented in radian units. The value of the fidelity, $  \mathcal{F}_{\text{Pr}_1}-\frac{1}{2}$,   $\mathcal{F}_{\text{Pr}_1}-\frac{2}{3}$, or $\mathcal{F}_{\text{Pr}_1}-0.789$ is indicated using color where the exact value corresponding to a particular color is demonstrated in the color box situated on the right side of the corresponding plot. The fidelities are dimensionless.}\label{fig2} 
\end{figure*}

In Fig.~\ref{fig1}, we plot the fidelity, $\mathcal{F}_{\text{Pr}_1}$, as a function of the control qubit's state's parameters, $\theta$ and $\phi$. To compare it with the values $\frac{1}{2}$, $\frac{2}{3}$, and 0.789, we also plot the planes $\mathcal{F}_{\text{Pr}_1}=\frac{1}{2}$,  $\mathcal{F}_{\text{Pr}_1}=\frac{2}{3}$, and $\mathcal{F}_{\text{Pr}_1}=0.789$ in the same figure. It can be understood from the figure that though $\mathcal{F}_{\text{Pr}_1}$ is not always greater than 0.789, $\frac{2}{3}$, or even $\frac{1}{2}$, there exists a wide range of values of $\theta$ and $\phi$ for which not only $\mathcal{F}_{\text{Pr}_1} > \frac{1}{2}$ or $\mathcal{F}_{\text{Pr}_1} > \frac{2}{3}$ but also $\mathcal{F}_{\text{Pr}_1} >0.789$. To understand precisely, the ranges of $\theta$ and $\phi$, for which $\mathcal{F}_{\text{Pr}_1}> \frac{1}{2}$, $\mathcal{F}_{\text{Pr}_1}>\frac{2}{3}$, or $\mathcal{F}_{\text{Pr}_1}>0.789$, in Fig.~\ref{fig2}, we plot $\mathcal{F}_{\text{Pr}_1}- \frac{1}{2}$ (left panel), $\mathcal{F}_{\text{Pr}_1}- \frac{2}{3}$ (middle panel), and  $\mathcal{F}_{\text{Pr}_1}- 0.789$ (right panel). The figure clearly indicates that though $\mathcal{F}_{\text{Pr}_1}$ is not always greater than 0.789 or $\frac{2}{3}$, it is almost always greater than $\frac{1}{2}$ except for two small regions of the ($\theta,\phi$)-plane. 

If instead of a control qubit, we consider the classical mixture of two situations, where with probability $|\alpha|^2$, the singlet is {being used}, and with probability $|\beta|^2$, it is not, then the average fidelity in absence of classical communication will be 
\begin{equation}
\mathcal{F}_{\text{Pr}_1}^{\text{mix}}= |\alpha|^2+\frac{|\beta|^2}{2}=\frac{1}{2}\left(1+\cos^2\left(\frac{\theta}{2}\right)\right). 
\end{equation}
This is equivalent to the case when the control qubit is present but is devoid of any quantum coherence in the $\{\ket{\text{on}},\ket{\text{off}}\}$ basis.
Since the classical mixture always provides fidelity greater than $\frac{1}{2}$ (fidelity corresponding to a random guess), whereas there are instances when fidelity using the quantum control is less than $\frac{1}{2}$, one may apparently conclude that the quantumness of the control does not have any overall advantage. But this is not the case. In the following subsection, we present a formal comparison between $\mathcal{F}_{\text{Pr}_1}$ and $\mathcal{F}_{\text{Pr}_1}^{\text{mix}}$.

\begin{figure*}
\includegraphics[scale=.50]{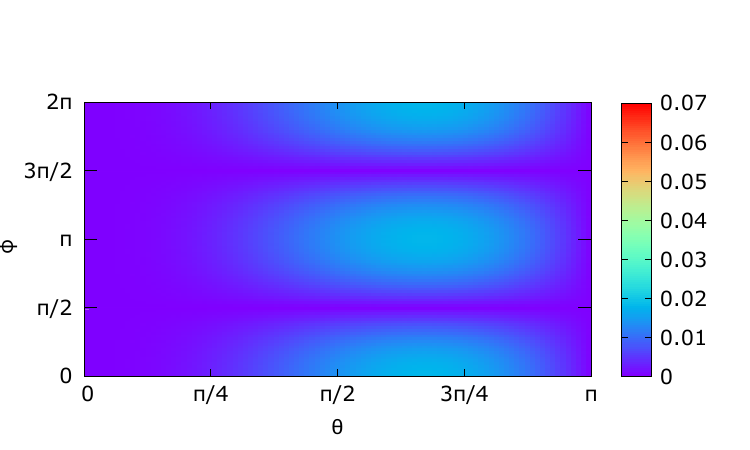}
\includegraphics[scale=.50]{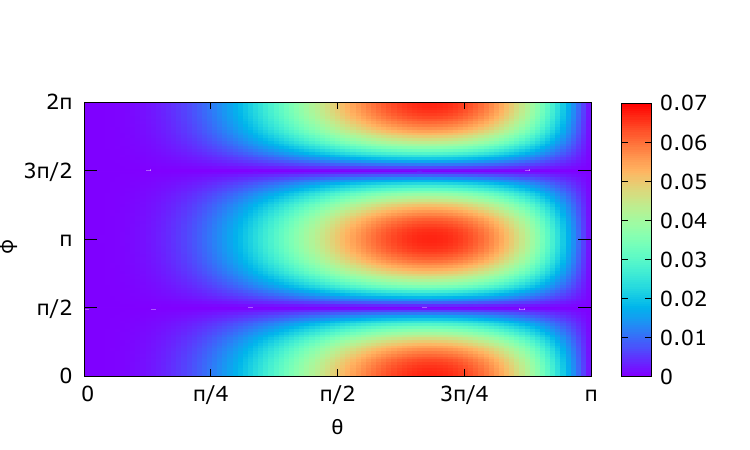}
\caption{Amount of quantum advantage. We plot the difference between the maximum fidelity, achieved by taking superposition (maximized over fidelities corresponding to the two scenarios, i.e., choosing $\ket{\text{on}}$ or $\ket{\text{off}}$, as a preferable measurement outcome) and the fidelity corresponding to taking classical mixture, of $\ket{\text{on}}$ and $\ket{\text{off}}$ states as initial state of the control, by performing Protocol 1 (left panel) and Protocol 2 (right panel). In particular, we plot $D_1^{max}$ and $D_2^{max}$ as functions of $\theta$ (horizontal axis) and $\phi$ (vertical axis) using colours. The angles $\theta$ and $\phi$ are considered in radian unit. The quantities represented using colours are dimensionless.
}\label{fig5} 
\end{figure*}

\subsection*{Do the quantumness provide any advantage?}
We measure the quantum advantage by examing the fidelity differences, that is,
\begin{equation}
    D_1(\theta,\phi)=\mathcal{F}_{\text{Pr}_1}-\mathcal{F}_{\text{Pr}_1}^{\text{mix}}=\frac{\sin\theta\cos\phi(\cos\theta-1)}{4(4-\sin\theta\cos\phi)}. \label{Fid1}
\end{equation}
If $D_1$ is positive, we conclude that the existent superposition in the control qubit provides an advantage over the unsuperposed control qubit state, which is a classical mixture of $\ket{\text{on}}$ and $\ket{\text{off}}$ with the corresponding probabilities. From Eq.~(\ref{Fid1}), it is clear that the positivity of $D_1$ depends only on the relative phase of the control qubit's initial state, i.e., the value of $\phi$. Whatever be the value of $\theta$, the superposed control qubit will always be more beneficial than the classical mixture for $\frac{\pi}{2}\leq\phi\leq\frac{3\pi}{2}$. But this range of $\phi$ will get altered to the complementary region if we choose $\ket{\text{off}}$ instead of $\ket{\text{on}}$ in the measurement performed on the control qubit at the end of the protocol. In that case, $\mathcal{F}'_{\text{Pr}_1}-\mathcal{F}_{\text{Pr}_1}^{\text{mix}}\geq 0$ is satisfied for $\phi\leq \frac{\pi}{2}$ and $\phi\geq \frac{3\pi}{2}$, where we have used the prime to denote the case when the outcome $\ket{\text{off}}$ is chosen in the measurement, instead of $\ket{\text{on}}$. Thus, if the freedom of choosing $\ket{\text{on}}$ or $\ket{\text{off}}$, depending on the initial state of the control qubit (in particular, of $\phi$), is provided, then the difference between the fidelities,
\begin{eqnarray}
    D_1^{max}(\theta,\phi)=\mathcal{F}^{max}_{\text{Pr}_1}-\mathcal{F}_{\text{Pr}_1}^{\text{mix}},
\end{eqnarray} 
become always non-negative. Here we have used the notation $\mathcal{F}^{max}_{\text{Pr}_1}=\max\{\mathcal{F}_{\text{Pr}_1},\mathcal{F}'_{\text{Pr}_1}\}$.
In Fig.~\ref{fig5}, we plot $D_1^{max}$ (in the left panel) as a function of $\theta$ and $\phi$. Hence, we see that Protocol 1 always provides advantage, except for a set of measure zero on the $(\theta,\phi)$-plane, in the sense that the fidelity of teleportation is always better than when a classical mixture of the $\ket{\text{on}}$ and $\ket{\text{off}}$ states of the quantum control is utilized.

Let us next consider the situation where classical communication is allowed between the two parties, {even when they do not use} any entanglement. And then we examine the extend of quantum advantage in using the control qubit.

\section{Role of classical communication}\label{sec4}
We are now going to introduce the second protocol. It's similar with the previous protocol, except that in this case, we have the facility of using classical communication {even when the singlet, shared between the parties, is not being used}. Note that it is reasonable to do so, as when the singlet is {being used}, the usual teleportation protocol is followed, which utilizes classical communication. Thus, in this protocol, when the singlet is not {being utilized} Alice measures her qubit \(|\psi\rangle_{A'}\) in an arbitrary basis (say, the computational basis), sends the information about the outcome classically to Bob, and Bob throws away his part of the singlet and prepares a qubit in a state that is same as the outcome at Alice. The corresponding transformation can mathematically be expressed as

\begin{eqnarray*}
\left(\left(a\ket{0}+b\ket{1}\right)_{A'}\otimes \ket{\psi^-}_{AB}\right)\left(\left(a^*\bra{0}+b^*\bra{1}\right)_{A'}\otimes  \bra{\psi^-}_{AB}\right)\longrightarrow\\
|a|^2\ket{0}_{A'}\bra{0}\otimes \left(\frac{\mathcal{I}_A}{2}\right)_{A}\otimes\ket{0}_{B}\bra{0}+|b|^2\ket{1}_{A'}\bra{1}\otimes \left(\frac{\mathcal{I}_A}{2}\right)_A\otimes\ket{1}_B\bra{1},\nonumber
\end{eqnarray*}
where $a\ket{0}+b\ket{1}$ represents the initial state $\ket{\psi}_{A'}$.
{The above transformation, though describes a classical event, can be expressed using the following Kraus operators}
\begin{eqnarray*}
   &&L_1=\ket{000}\bra{001}\text{, }L_2=\ket{010}\bra{010}\text{, }\\ &&L_3=\ket{101}\bra{101}\text{, }L_4=\ket{111}\bra{110}.
\end{eqnarray*}
It can be easily checked that $\sum_\nu L_\nu (\ket{\psi}\bra{\psi})_{A'}(\ket{\psi^-}\bra{\psi^-})_{AB} L_\nu^\dagger=|a|^2\ket{0}_{A'}\bra{0}\otimes \left(\frac{\mathcal{I}_A}{2}\right)_{A}\otimes\ket{0}_{B}\bra{0}+|b|^2\ket{1}_{A'}\bra{1}\otimes \left(\frac{\mathcal{I}_A}{2}\right)_A\otimes\ket{1}_B\bra{1}. $ But the set of operators, $\{L_\nu\}_\nu$, does not satisfy the completeness relation, that is $\sum_{\nu=1}^4 L_\nu^\dagger L_\nu \neq \mathcal{I}_{AA'B}$. To complete the set of Kraus operators, we additionally define the four Kraus operators,
\begin{eqnarray*}
    &&L_5=\ket{\chi_1}\bra{000}\text{, }L_6=\ket{\chi_2}\bra{011}\text{, }\\ &&L_7=\ket{\chi_3}\bra{100}\text{, }L_8=\ket{\chi_4}\bra{111}.
\end{eqnarray*}
For an arbitrary normalized set of states $\{\ket{\chi_i}\}_i$, the operators satisfy $\sum_{\nu=1}^8 L_\nu^\dagger L_\nu=\mathcal{I}_{A'AB}$ and $L_\nu (\ket{\psi}\bra{\psi})_{A'}(\ket{\psi^-}\bra{\psi^-})_{AB} L_\nu^\dagger=0$ for $\nu=$5, 6, 7, 8.

To describe the entire process of \textbf{Protocol 2}, we define the set of Kraus operators, 
\begin{equation*}
    N_{\mu\nu}=\frac{1}{\sqrt{8}}\ket{\text{on}}\bra{\text{on}}\otimes K_\mu+\frac{1}{2}\ket{\text{off}}\bra{\text{off}}\otimes L_\nu,
\end{equation*}
which acts on the composite Hilbert space consisting of $C$, $A$, $A'$, and $B$. Hence the final output state of the protocol is 
    \begin{eqnarray*}
   \rho^{\text{Pr}_2}_{CA'AB}= \sum_{\mu=1}^4\sum_{\nu=1}^8 N_{\mu\nu} \ket{\xi}\bra{\xi}
   N_{\mu\nu}^\dagger.
\end{eqnarray*}
Similar to the previous protocol, now we apply the Hadamard gate on the control qubit's state and measure the control qubit in the basis $\{\ket{\text{on}},\ket{\text{off}}\}$. The state corresponding to the output $\ket{\text{on}}$ is
\begin{eqnarray*}
\rho^{\text{Pr}_2}_{A'AB}
   =\frac{|\alpha|^2}{2}\sum_{\mu=1}^4 K_\mu\rho K_\mu^\dagger+\frac{\alpha\beta^*}{4\sqrt{8}}\left(\sum_{\mu=1}^4 K_\mu\right)\rho\left(\sum_{\nu=1}^8 L_\nu^\dagger\right)+\\ \frac{\alpha^*\beta}{4\sqrt{8}}\left(\sum_{\nu=1}^8 L_\nu\right)\rho\left(\sum_{\mu=1}^4 K_\mu^\dagger\right)+\frac{|\beta|^2}{2}\sum_{\nu=1}^8 L_\nu\rho L_\nu^\dagger.
\end{eqnarray*}
The above state is not normalized. After normalizing and tracing out $A'$ and $A$, we are left with Bob's system described by the following state:
\begin{eqnarray*}
\rho^{\text{Pr}_2}_{B}&=&\frac{16|\alpha|^2-\sqrt{2}(\alpha^*\beta+\beta^*\alpha)}{16-\sqrt{2}(\alpha^*\beta+\beta^*\alpha)}\ket{\psi}\bra{\psi}\\
    &&+\frac{16|\beta|^2}{16-\sqrt{2}(\alpha^*\beta+\beta^*\alpha)}[|a|^2\ket{0}\bra{0}+|b|^2\ket{1}\bra{1}].
\end{eqnarray*}
Fidelity of the state, $\rho^{\text{Pr}_2}_{B}$, to the expected state $\ket{\psi}\bra{\psi}$ is 
\begin{eqnarray*}
F_1&=&\frac{16|\alpha|^2-\sqrt{2}(\alpha^*\beta+\beta^*\alpha)+16|\beta|^2(|a|^4+|b|^4)}{16-\sqrt{2}(\alpha^*\beta+\beta^*\alpha)}\\
    &=&\frac{16|\alpha|^2-\sqrt{2}(\alpha^*\beta+\beta^*\alpha)+16|\beta|^2 \left(\cos^4{\frac{\theta'}{2}}+\sin^4{\frac{\theta'}{2}}\right) }{16-\sqrt{2}(\alpha^*\beta+\beta^*\alpha)}
\end{eqnarray*}
Here we have set $a =\cos(\theta'/2)$ and $b = e^{i\phi'}\sin(\theta'/2)$, where $\theta' \in [0, \pi]$ and $\phi' \in [0, 2\pi)$. Since the fidelity, $F_1$, depends on
the initial state of the qubit $A$, to have an overall idea about
the success of the protocol, we take average of $F_1$ over the
complete set of input states. We get
\begin{eqnarray}
\mathcal{F}_{\text{Pr}_2}&=&\frac{1}{4\pi}\int_0^{2\pi}\int_0^\pi F_1 \sin\theta' d\theta' d\phi'\nonumber\\
&=&\frac{40+8\cos \theta-3\sqrt{2}\sin\theta\cos\phi}{48-3\sqrt{2}\sin\theta\cos\phi}. \label{eq3}
\end{eqnarray}

\begin{figure}[b!]
\includegraphics[scale=1.0]{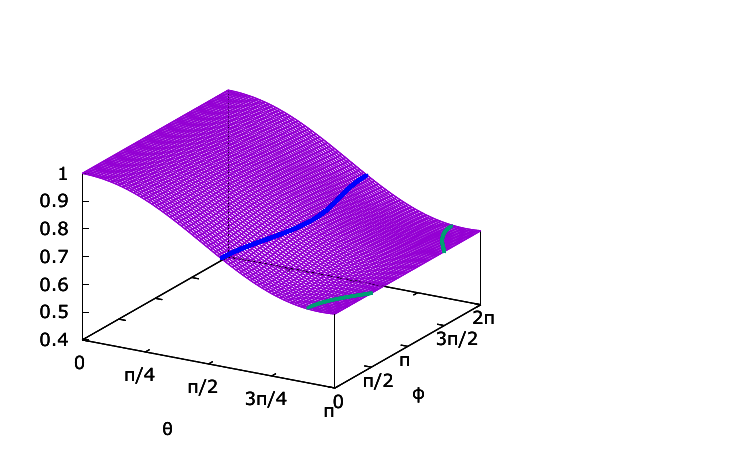}
\caption{Performance of the second teleportation protocol (Protocol 2). We plot here the fidelity of the output state of Bob's qubit to the actual expected state after going through Protocol 2. $\mathcal{F}_{\text{Pr}_2}$ [see Eq.~(\ref{eq3})] is plotted along the vertical axis against $\theta$ and $\phi$ along the horizontal axes, using radian as the unit. The vertical axis is dimensionless. To compare the fidelity with the classical case, we also plot the curve along which the planes $\mathcal{F}_{\text{Pr}_2}=\frac{2}{3}$ (classically attainable fidelity with the measure-prepare process) and $\mathcal{F}_{\text{Pr}_2}=0.789$ (the bound on fidelity beyond which the protocol shows genuine quantum nature)
cuts the $\mathcal{F}_{\text{Pr}_2}(\theta, \phi)$ surface using green and blue lines, respectively, in the same graph.}\label{fig3} 
\end{figure}

\begin{figure*}
\includegraphics[scale=.50]{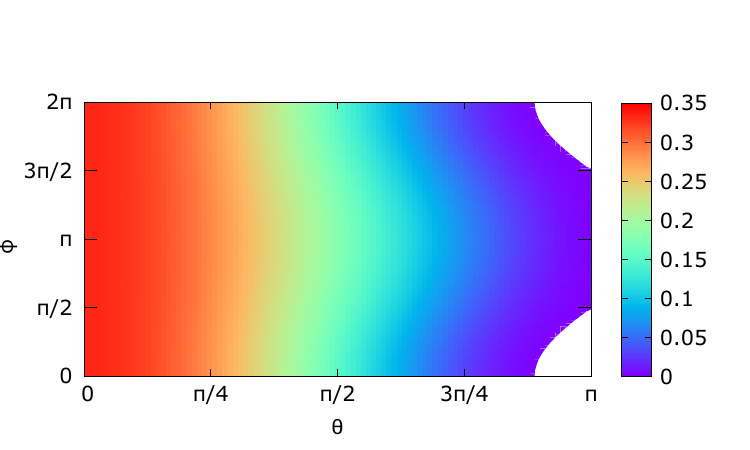}
\includegraphics[scale=.50]{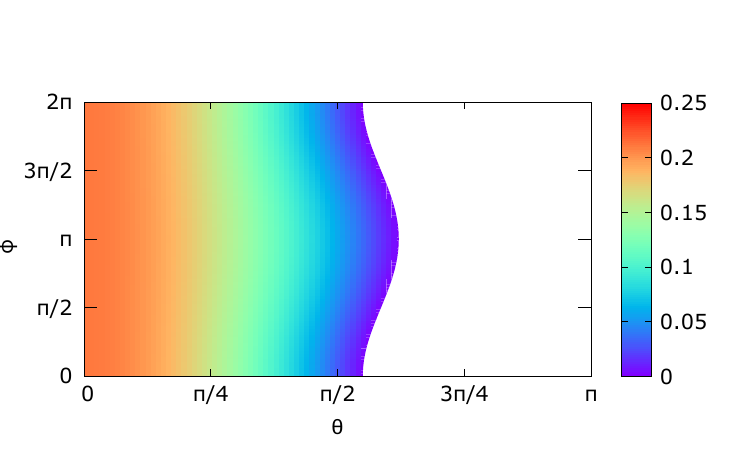}
\caption{Comparison between the fidelity attained using Protocol 2 and the maximum attainable fidelity using classical method. All considerations here are the same as in the middle and right panels of Fig. \ref{fig2}, except the fact that here we have considered Protocol 2, and thus, instead of plotting $\mathcal{F}_{\text{Pr}_1}-\frac{2}{3}$ and $\mathcal{F}_{\text{Pr}_1}-0.789$, we plot $\mathcal{F}_{\text{Pr}_2}-\frac{2}{3}$ (left panel) and $\mathcal{F}_{\text{Pr}_2}-0.789$ (right panel).}\label{fig4} 
\end{figure*}

It is easy to check that the above fidelity is greater than $\frac{1}{2}$. In Fig.~\ref{fig3}, we plot $\mathcal{F}_{\text{Pr}_2}$ with respect to $\theta$ and $\phi$. To compare it with the maximum classical fidelity in the measure-prepare protocol and the maximum classical fidelity achievable in the absence of any genuine quantum resource, we also plot the curves at which the planes $\mathcal{F}_{\text{Pr}_2}=\frac{2}{3}$ and $\mathcal{F}_{\text{Pr}_2}=0.789$ cut the surface defined in (\ref{eq3}), in the same figure. To more precisely indicate the ranges of values of the parameters $\theta$ and $\phi$ for which $\mathcal{F}_{\text{Pr}_2}$ outperforms classical fidelities, in Fig.~\ref{fig4}, we exhibit projected plots of $\mathcal{F}_{\text{Pr}_2}-\frac{2}{3}$ and $\mathcal{F}_{\text{Pr}_2}-0.789$. 
Again, if we consider the situation where {having the authority to use the singlet} depends on a probability distribution, i.e., with $|\alpha|^2$ probability the singlet is {being used} and with probability $|\beta|^2$ it is not, the corresponding fidelity will be 
\begin{equation}
      F_2= |\alpha|^2+|\beta|^2(|a|^4+|b|^4)=|\alpha|^2+|\beta|^2\left(\cos^4{\frac{\theta'}{2}}+\sin^4{\frac{\theta'}{2}}\right).
\end{equation}
Hence the average fidelity, with the average being taken over all possible initial pure states of $A'$, is given by
\begin{eqnarray}
\mathcal{F}^{\text{mix}}_{\text{Pr}_2} =\frac{1}{4\pi}\int_0^{2\pi}\int_0^\pi F_2 \sin\theta' d\theta' d\phi'
=\frac{1}{3}\left(2+\cos^2 \frac{\theta}{2}\right)
\end{eqnarray}
Clearly, the average fidelity in this case is $\geq\frac{2}{3}$ for all $\theta$ and $\phi$. 

In the following subsection, we discuss about the difference between the effectivenesses of the second protocol using quantum control and using a classical mixture and examine the regions of the parameter space where the quantum control provides advantage.

\subsection*{Comparison between the quantum control and classical mixture}
We have considered two cases corresponding to Protocol 2, viz. the control qubit has quantum coherence and it is a classical mixture. The difference between the two corresponding average fideliteis is given by
\begin{eqnarray}
D_2(\theta,\phi)=\mathcal{F}_{\text{Pr}_2}-\mathcal{F}_{\text{Pr}_2}^{\text{mix}}=\frac{\sin\theta\cos\phi(\cos\theta-1)}{\sqrt{2}\left(48-3\sqrt{2}\sin\theta\cos\phi\right)}.\nonumber\\ \label{fid2}
\end{eqnarray}
It is evident from the Eq.~(\ref{fid2}) that $D_2(\theta,\phi)\geq 0$ for $\frac{\pi}{2}\leq\phi\leq\frac{3\pi}{2}$, i.e., in this region, quantum advantage can be achieved. 

If we consider the scenario where after performing the measurement on the control qubit, the measurement output $\ket{\text{off}}$ is being chosen instead of $\ket{\text{on}}$, the corresponding fidelity of Bob's state would be $\mathcal{F}'_{\text{Pr}_2}(\theta,\phi)    =\frac{40+8\cos \theta+3\sqrt{2}\sin\theta\cos\phi}{48+3\sqrt{2}\sin\theta\cos\phi}$, so that $\mathcal{F}'_{\text{Pr}_2}(\theta,\phi)=\mathcal{F}_{\text{Pr}_2}(\theta,\phi+\pi)$. By choosing between $\mathcal{F}_{\text{Pr}_2}$ and $\mathcal{F}'_{\text{Pr}_2}$, we can cover the full range of $\phi$ such that quantum advantage is inevitable. In Fig.~\ref{fig5}, we plot
\begin{equation}
D_2^{max}=\max\{\mathcal{F}'_{\text{Pr}_2},\mathcal{F}_{\text{Pr}_2}\}-\mathcal{F}_{\text{Pr}_2}^{\text{mix}}\label{eqfin2}
\end{equation}
in the right panel as a function of $\theta$ and $\phi$. From the figure it is evident that the quantum advantage in Protocol 1 is much more intense than in Protocol 2.

\section{Quantum advantage and Quantum coherence of control qubit's state}\label{sec5}
From Eqs.~(\ref{Fid1}) and (\ref{fid2}), it is clear that the amount of quantum advantage  depends on the control qubit's initial state, in particular, the value of $\alpha$ and $\beta$. Since the initial state of control qubit, $|\alpha|^2\ket{\text{on}}\bra{\text{on}}+|\beta|^2\ket{\text{off}}\bra{\text{off}}$, would not have given any quantum advantage (when the reference basis is $\{\ket{\text{on}}, \ket{\text{off}}\}$), we expect that the resource behind the quantum advantage is provided by the superposition in the control qubit's initial state. Therefore, we want to explore the relation between the quantum coherence of the control qubit's state, which is a quantifier of superposition, and the quantum advantage, in the context of the two protocols individually. 

The $l_1$-norm of quantum coherence, $C$, of a state, say $\sigma$, in an arbitrary but fixed basis $\{\ket{i}\}_i$, is given by~\cite{coh1, coh2, coh3}
\begin{equation*}
    C(\sigma)=\sum_{i,j,i\neq j} |\bra{i}\sigma\ket{j}|.
\end{equation*}
 Thus, we see that the quantum coherence of a state is a basis-dependent quantity. For more details on quantum coherence readers can go through the review papers \cite{coh_rev1,coh_rev2}.

The initial state of the control qubit is considered to be $\alpha\ket{\text{on}}+\beta\ket{\text{off}}$. We assume that $\{\ket{\text{on}}$, $\ket{\text{off}}\}$ are eigenstates corresponding to the Pauli matrix $\sigma_z$. Therefore, the quantum coherence of the control qubit's state in $\sigma_z$ basis, i.e., with respect to $\{\ket{\text{on}},\ket{\text{off}}\}$, is 
\begin{equation*}
    c_z=|\alpha^*\beta|+|\alpha\beta^*|=\sin{\theta},
\end{equation*}
whereas the coherence in $\sigma_x$ basis, i.e., with respect to $\Big\{\frac{\ket{\text{on}}+\ket{\text{off}}}{\sqrt{2}},\frac{\ket{\text{on}}-\ket{\text{off}}}{\sqrt{2}}\Big\}$, is given by
\begin{eqnarray*}
    c_x&=&\frac{1}{2}(|\cos{\theta}+i\sin{\theta}\sin{\phi}|+|\cos{\theta}-i\sin{\theta}\sin{\phi}|)\\ &=&\sqrt{1-\sin^2\theta\cos^2\phi}.
\end{eqnarray*}
The amount of quantum advantage provided in Protocol 1 can be expressed in terms of the coherence, $c_z$, in the following way:
\begin{eqnarray*}
    \Delta_1(c_z,\phi)&=& \frac{c_z \cos\phi \left(\sqrt{1-c_z^2}-1\right)}{4(4-c_z\cos\phi)} \text{ for }0\leq \theta \leq \frac{\pi}{2},\\
   &=&-\frac{c_z \cos\phi \left(\sqrt{1-c_z^2}+1\right)}{4(4-c_z\cos\phi)} \text{ for }\frac{\pi}{2}\leq \theta \leq \pi.
\end{eqnarray*}
Let us first focus on the first range, i.e., $0\leq \theta \leq \frac{\pi}{2}$. The corresponding function, $\Delta_1(c_z,\phi)$, is clearly positive in the range $\frac{\pi}{2}\leq \phi \leq \frac{3\pi}{2}$. 

The derivative of $\Delta_1(c_z,\phi)$ with respect to $c_z$ is given by
\begin{eqnarray*}
   \frac{\partial\Delta_1(c_z,\phi)}{\partial c_z}= \frac{\cos\phi\left(4-4\sqrt{1-c_z^2}-8c_z^2+c_z^3\cos\phi\right)}{4(4-c_z\cos\phi)^2\sqrt{1-c_z^2}}\nonumber\\\text{ for }0\leq \theta \leq \frac{\pi}{2}\nonumber.
\end{eqnarray*}
It can be easily checked that the function $4-4\sqrt{1-c_z^2}-8c_z^2$ has maximum at $c_z=0$ and the corresponding maximum value is 0. Moreover, since  $c_z\geq 0$ and $\cos\phi\leq0$ for $\frac{\pi}{2}\leq \phi \leq \frac{3\pi}{2}$, $4-4\sqrt{1-c_z^2}-8c_z^2+c_z^3 \cos\phi\leq c_z^3 \cos\phi\leq 0$, which implies $\frac{\partial\Delta_1(c_z,\phi)}{\partial c_z}\geq 0$. Hence, we conclude within the range $0\leq \theta\leq\frac{\pi}{2}$ when $\Delta_1(c_z,\phi)$ is positive, the function is a monotonically increasing function of $c_z$, that is, the advantage of using Hadamard gate increases with the coherence in the initial control qubit's state.

Let us now move to the second range, i.e., $\frac{\pi}{2}\leq \theta \leq \pi$. It is surely negative within the range $\phi\leq\frac{\pi}{2}$ or $\phi\geq\frac{3\pi}{2}$ and positive for $\frac{\pi}{2}\leq\phi\leq\frac{3\pi}{2}$. Since we want to examine the situation where the utilization of quantum control is advantageous, we are interested in the second region. The derivative of the function, $\Delta_1(c_z,\phi)$, with respect to $c_z$ is given by
\begin{eqnarray*}
    \frac{\partial \Delta_1(c_z,\phi)}{\partial c_z}=-\frac{\cos\phi\left(4+4\sqrt{1-c_z^2} -8 c_z^2+c_z^3\cos\phi\right)}{4(4-c_z\cos\phi)^2\sqrt{1-c_z^2}}\nonumber\\\text{ for }\frac{\pi}{2}\leq \theta \leq \pi.
\end{eqnarray*}
The above function can be both positive and negative, depending on the particular values of $c_z$ and $\phi$. For example, $\frac{\partial \Delta_1(0,\pi)}{\partial c_z}=0.125>0$ and $\frac{\partial \Delta_1(0.9,\pi)}{\partial c_z}=-0.035<0$. Therefore, in this range, the quantum coherence of control qubit cannot guarantee the increasing quantum advantage. We have expressed the quantum advantage, i.e., $D_1(\theta,\phi)$ or $\Delta_1(c_z,\phi)$, in terms of the quantum coherences $c_z$ and $c_x$ in the Appendix. 

Let us now move to Protocol 2 where classical communication is allowed even when the control qubit is in the $\ket{\text{off}}$ state. The quantum advantage in this protocol can also be written in terms of the coherence $c_z$ and can be expressed as
\begin{eqnarray*}
    \Delta_2(c_z,\phi)&=&\frac{c_z \cos\phi \left(\sqrt{1-c_z^2}-1\right)}{\sqrt{2}\left(48-3\sqrt{2}c_z\cos\phi\right)}\text{ for }0\leq\theta\leq\frac{\pi}{2},\\
    &=&-\frac{c_z \cos\phi \left(\sqrt{1-c_z^2}+1\right)}{\sqrt{2}\left(48-3\sqrt{2}c_z\cos\phi\right)}\text{ for }\frac{\pi}{2}\leq\theta\leq\pi.
\end{eqnarray*}
Considering each range separately and following the same path of logic and calculations as in the previous one, we obtain the following points:
\begin{enumerate}
    \item 
    Within the range $0\leq\theta\leq\frac{\pi}{2}$ and $\phi\leq\frac{\pi}{2}$ or $\phi\geq\frac{3\pi}{2}$, using classical mixture is more advantegeous than superposed state of control qubit.
    \item
    Quantum advantage can be gained within the range $0\leq\theta\leq\frac{\pi}{2}$,  $\frac{\pi}{2}\leq\phi\leq\frac{3\pi}{2}$. Moreover, the amount of quantum advantage is a monotonically increasing function of the quantum coherence, $c_z$.
    \item
    In the range $\frac{\pi}{2}\leq\theta\leq\pi$ and $\phi\leq\frac{\pi}{2}$ or $\phi\geq\frac{3\pi}{2}$, again there will not be any advantage from quantum superposition.
    \item
    In the remaining region, i.e., $\frac{\pi}{2}\leq\theta\leq\pi$ and $\frac{\pi}{2}\leq\phi\leq\frac{3\pi}{2}$, though one achieves quantum advantage, increasing quantum coherence may not help to raise the advantage.
\end{enumerate}
As we have discussed earlier in detail, the above points are also true for Protocol 1.

The difference in the fidelities of using superposition and classical mixture in Protocol 2, $D_2(\theta,\phi)$ or $\Delta_2(c_z,\phi)$, can be represented as a function of the two coherences, $c_z$ and $c_x$. One can go through the Appendix for the exact expressions. 

\section{Conclusion}\label{sec6}
We considered the superposition of two situations of teleporting a quantum bit, one where an entanglement resource is allowed to be used and another where it is available but not permitted to be used. Setting up the scenario requires the inclusion of an additional two-level system that controls if Alice and Bob, the sender and receiver respectively of the teleportation protocol, are authorized to use a shared maximally entangled state. When this control qubit is ``on", Alice and Bob use a shared maximally entangled state, and thus the usual teleportation protocol can be performed, whereas when they do not use any shared entanglement, that is, when the control qubit is in the ``off" state, Alice and Bob can either do nothing (Protocol 1) or use the classical communication (Protocol 2) to teleport the state with imperfect fidelity. By teleporting via classical communication, we mean that Alice measures the state which she wants to teleport on an arbitrary but fixed basis, communicates the measurement outcome to Bob, and Bob prepares the output state in his lab. The control qubit {that dictates if the shared entanglement will be used or not} was initially taken to be a product with the remaining part of the set-up. Through the protocol, entanglement can be generated between the control qubit and the rest of the system, but we are not creating any additional entanglement between Alice and Bob.
After accomplishing the teleportation protocol, we applied a rotation on the control qubit's state, measured it, and considered the output corresponding to a particular state. 

We compared the average fidelity of the output state to the initial target state with the average fidelity of a random guess and with the maximum average fidelity attainable using classical communication. We see that Protocol 2 always gives a fidelity that is greater than a random guess, whereas in Protocol 1, there are situations where the fidelity can be lower than the fidelity corresponding to the random guess. This is intuitively satisfactory, since in Protocol 2 we have used classical communication when entanglement is absent.

An obvious question arises here: can there be any quantum advantage due to the quantum coherence in the control qubit? We see for both protocols, the answer depends only on the phase of the superposition and not on the probability amplitudes. We examine if there exists any relation between the amount of quantum advantage and quantum coherence in the control qubit's initial state. The connection between the quantum coherence and the difference between the fidelities corresponding to using superposed state of the control qubit and classical mixture depends on the range of the parameters defining the superposition of $\ket{\text{on}}$ and $\ket{\text{off}}$ states of the control qubit. In particular, for a particular range, the difference is a monotonically increasing function of the quantum coherence.

A fundamental question in quantum information theory is whether nonclassicality can be detected in the output state of a physical process. This question can be naturally generalized in the following way: if we consider the superposed situation of two events, of which one is classical and the other nonclassical, then to what extend is the output nonclassical? We have tried to address this in the context of quantum teleportation. Within this study, we have explored a hitherto unexplored role of superposition in quantum teleportation.  

Considering such a superposition, we show that there is a quantum advantage in the sense that if a classical mixture had been considered, the corresponding fidelity of the received state to the target state would be less than that which can be achieved by superposing the events. Though we do not know any practical scenario where this kind of situation may arise where a control qubit can dictate if the sender and the receiver performing the teleportation protocol are allowed to use the singlet or not, we hope that if somehow, in the future, such a superposition can be created in physical systems, that can provide an advantage over taking a classical mixture. We hope experiments in this direction will make the subject more transparent. 

\section*{Acknowledgments}
The research of KS was supported in part by the INFOSYS scholarship. We acknowledge partial support from the Department of Science and Technology, Government
of India through the QuEST grant (grant number
DST/ICPS/QUST/Theme3/2019/120).

\section*{APPENDIX}
$D_1(\theta,\phi)$ or $\Delta_1(c_z,\phi)$, can be expressed in terms of the coherences $c_z$ and $c_x$. The functional form depends on the range of $\theta$ and $\phi$. Considering the four distinct ranges of the parameters individually, we have presented the functional forms of $\mathcal{G}_1(c_z,c_x)=\mathcal{F}_{\text{Pr}_1}-\mathcal{F}_{\text{Pr}_1}^{\text{mix}}$ in the following way: (i) For $0\leq\theta\leq\frac{\pi}{2}$ and $\phi\leq\frac{\pi}{2}$ or $\phi\geq\frac{3\pi}{2}$,
\begin{equation*}
\mathcal{G}_1(c_z,c_x)=\frac{\sqrt{1-c_x^2}\left(\sqrt{1-c_z^2}-1\right)}{4\left(4-\sqrt{1-c_x^2}\right)}.
\end{equation*}
(ii) For $0\leq\theta\leq\frac{\pi}{2}$ and $\frac{\pi}{2}\leq \phi\leq\frac{3\pi}{2}$,
\begin{equation*}
\mathcal{G}_1(c_z,c_x)=-\frac{\sqrt{1-c_x^2}\left(\sqrt{1-c_z^2}-1\right)}{4\left(4+\sqrt{1-c_x^2}\right)}.
\end{equation*}
(iii) For $\frac{\pi}{2}\leq\theta\leq\pi$ and $\phi\leq\frac{\pi}{2}$ or $\phi\geq\frac{3\pi}{2}$,
\begin{equation*}
\mathcal{G}_1(c_z,c_x)=-\frac{\sqrt{1-c_x^2}\left(\sqrt{1-c_z^2}+1\right)}{4\left(4-\sqrt{1-c_x^2}\right)}.
\end{equation*}
(iv) For $\frac{\pi}{2}\leq\theta\leq\pi$ and $\frac{\pi}{2}\leq\phi\leq\frac{3\pi}{2}$,
\begin{equation*}
\mathcal{G}_1(c_z,c_x)=\frac{\sqrt{1-c_x^2}\left(\sqrt{1-c_z^2}+1\right)}{4\left(4+\sqrt{1-c_x^2}\right)}.
\end{equation*}

In case of Protocol 2, $D_2(\theta,\phi)$ or $\Delta_2(c_z,\phi)$, can also be represented as a function of the two coherences, $c_z$ and $c_x$ in the following way: (i) For $0\leq\theta\leq\frac{\pi}{2}$ and $\phi\leq\frac{\pi}{2}$ or $\phi\geq\frac{3\pi}{2}$,
\begin{equation*}
\mathcal{G}_2(c_z,c_x)=\frac{\sqrt{1-c_x^2}\left(\sqrt{1-c_z^2}-1\right)}{\sqrt{2}\left(48-3\sqrt{2}\sqrt{1-c_x^2}\right)}.
\end{equation*}
(ii) For $0\leq\theta\leq\frac{\pi}{2}$ and $\frac{\pi}{2}\leq \phi\leq\frac{3\pi}{2}$,
\begin{equation*}
\mathcal{G}_2(c_z,c_x)=-\frac{\sqrt{1-c_x^2}\left(\sqrt{1-c_z^2}-1\right)}{\sqrt{2}\left(48+3\sqrt{2}\sqrt{1-c_x^2}\right)}.
\end{equation*}
(iii) For $\frac{\pi}{2}\leq\theta\leq\pi$ and $\phi\leq\frac{\pi}{2}$ or $\phi\geq\frac{3\pi}{2}$, 
\begin{equation*}
\mathcal{G}_2(c_z,c_x)=-\frac{\sqrt{1-c_x^2}\left(\sqrt{1-c_z^2}+1\right)}{\sqrt{2}\left(48-3\sqrt{2}\sqrt{1-c_x^2}\right)}.
\end{equation*}
(iv) For $\frac{\pi}{2}\leq\theta\leq\pi$ and $\frac{\pi}{2}\leq\phi\leq\frac{3\pi}{2}$, 
\begin{equation*}
\mathcal{G}_2(c_z,c_x)=\frac{\sqrt{1-c_x^2}\left(\sqrt{1-c_z^2}+1\right)}{\sqrt{2}\left(48+3\sqrt{2}\sqrt{1-c_x^2}\right)}.
\end{equation*}

\section*{Reference}
\bibliographystyle{iopart-num}

\end{document}